\documentstyle[prb,twocolumn,aps]{revtex}
\begin{document}
\draft

\tolerance 50000

\twocolumn[\hsize\textwidth\columnwidth\hsize\csname@twocolumnfalse\endcsname

\title{ Conformal Field Theory and the Exact Solution of the 
BCS Hamiltonian}

\author{Germ{\'a}n  Sierra  
} 
\address{
Instituto de Matem{\'a}ticas y F{\'\i}sica Fundamental, C.S.I.C.,
Madrid, Spain. 
}

\maketitle 

\begin{abstract} 
\begin{center}
\parbox{14cm}{We propose a connection  between conformal field
theory (CFT) and 
the exact solution
and integrability of the reduced BCS model of superconductivity. 
The relevant CFT is given
by the $SU(2)_k$-WZW model in the singular limit when the level
$k$ goes to $-2$. This theory has to be perturbed by an operator
proportional to the inverse of the BCS coupling constant. 
Using the free field realization 
of this perturbed Wess-Zumino-Witten  model, 
we derive  the exact
Richardson's wave function  and the integrals of motion of the
reduced BCS model in the saddle point approximation.  
The construction is reminiscent of the CFT approach
to the Fractional Quantum Hall effect. 
}

\end{center}
\end{abstract}

\pacs{
\hspace{2.5cm} 
}
\vskip2pc] \narrowtext

\section*{I) Introduction}

The BCS theory has been used for decades
to describe the superconducting properties of 
``low $T_c$'' metallic materials\cite{BCS}. 
The starting point of the
theory is a Hamiltonian which describes 
the attractive interaction between the electrons in well
defined energy levels. The grand canonical 
BCS wave function gives a very
accurate solution of the BCS Hamiltonian in the limit
where the number of electrons, $N_e$, is very large.
The BCS wave function may also be projected to a fixed
number of electrons, $N_e$, giving essentially the same
physics when $N_e >>1$.   
However for small values
of $N_e$,  one has to use exact analytical or 
numerical methods
to obtain reliable results. 
The study of small fixed-$N_e$ superconductivity has a long story
which goes back to an old question posed by Anderson
as to what is the smaller size of a metallic
particle to remain  superconducting \cite{A}. 
Recent experiments \cite{RBT} involving 
aluminium grains with  nanometer size have 
inspired a number of theoretical works
where Anderson's question is reconsidered 
\cite{vD,Braun1,BvD,SA,ML,MFF,BH,Braun2,DS,DP,BD,S}.

In this paper we shall be concerned with the exact
analytic solution of the reduced BCS Hamiltonian
proposed by Richardson 
in a series of papers 
between 1963 and 1977 \cite{R1,R1bis,RS,R2,R3,R4,R5,R6,R7}. 
 This solution emerged in the framework
of Nuclear Physics and has passed unnoticed by most of the physics
comunity until the recent upheaval in ultrasmall metallic grains. 
A closely related work is that of Cambiaggio, Rivas and Saraceno
(CRS) who proved recently  the integrability of
the reduced BCS Hamiltonian without recourse to   
Richardson's solution \cite{CRS}. These authors
found a set of integrals of motion whose number equals that 
of the degrees of freedom of the system.   
Our aim is to show that Richardson's solution, 
together with its integrability, can be naturally  understood
in the framework  of conformal field theory (CFT). 
In other words, 
we shall propose a correspondence
between CFT and the BCS theory which gives a neat picture  
of the Richardson's wave function and the conserved quantities
found by CRS, which are in that sense  unified in a common framework. 
Our work is reminiscent 
to the application  of CFT 
to the  Laughlin wave function \cite{L} of the Fractional 
Quantum Hall effect (FQHE) \cite{F}.

The organization of the paper is as follows.
In sections II and III we present brief reviews
on the exact solution of the reduced BCS Hamiltonian
and the free field realization of the $SU(2)_k$-WZW model.
 In section IV we derive
the exact Richardson's solution and we show integrability
of the BCS model  
using the free field realization of  the $SU(2)_k$-WZW model.
In section V we explain the BCS/CFT connection in a second
quantized language. In section VI we explore the analogies
and differences between the CFT approaches to  BCS
and the FQHE. 
Finally, in section VII we state  our conclusions and
prospects of future work.

\section*{II) Review of the exact solution of 
the BCS model}

The reduced BCS model is defined by the Hamiltonian 
\cite{BCS,vD,Braun1,BvD,SA,ML,MFF,BH,Braun2,DS,DP,BD,S}

\begin{eqnarray}
H_{BCS} = \sum_{j, \sigma= \pm} 
\varepsilon_{j\sigma} c_{j \sigma}^\dagger c_{j \sigma}
  -g d \sum_{j, j'}  c_{j +}^\dagger c_{j -}^\dagger 
c_{j' -} c_{j' +} \; 
\label{1}
\end{eqnarray}

\noindent where $c_{j,\pm}$ ( resp. $c^\dagger_{j,\pm}$)
is an electron  
destruction ( resp. creation) operator   
in the time-reversed states $|j, \pm \rangle$
with energies $\varepsilon_j$, $d$ is the  mean level spacing  and
$g$ is the BCS dimensionless coupling constant.
The sums in (\ref{1}) run over a set of $\Omega$ doubly
degenerate energy levels $\varepsilon_j ( j=1,\dots, \Omega)$. 
We shall assume in this paper that the energy levels  are all
distinct, i.e. $\varepsilon_i \neq \varepsilon_j $ for $i \neq j$.  
The Hamiltonian (\ref{1}) is really a simplified 
version of the reduced BCS Hamiltonian where all couplings
have been  set equal to a single one,  namely $g$. 
This is the model that is commoly used to described
ultrasmall grains, which describes the scattering 
of pairs of electrons between discrete energy levels
that come in time-reversed states. 
Hereafter  we shall refer to  
(\ref{1}) simply as  the BCS Hamiltonian.

As mentioned in the Introduction,   
Richardson had long ago solved this model exactly for 
an arbitrary set of levels, $\varepsilon_j$,  not necessarily
all distinct  \cite{R1,R1bis,RS,R2,R3,R4,R5,R6,R7}. 
To simplify matters,  we shall assume that
there are not singly occupied electronic levels. 
As can be seen from (\ref{1}), these levels decouple
from the rest of the  system;  they are said to be blocked,  
contributing only  with their energy  
$\varepsilon_j$ to the total energy $\cal E$. 
The above simplification implies that every energy level
$j$ is either empty ( i.e. $| {\rm vac}\rangle$),
 or occupied by a pair of electrons ( i.e.
$c^\dagger_{j,+} c^\dagger_{j,-} |{\rm vac}\rangle$). 
Denote the total number of electrons pairs by 
$N$. Then  of course
$N \leq \Omega$. The most studied case
in the literature corresponds to the half-filled situation,  
where the number of electrons,  $N_e= 2N$, is  equal to the number of levels
$\Omega$ 
\cite{vD,Braun1,BvD,SA,ML,MFF,BH,Braun2,DS,DP,BD,S}. 
In the absence of interaction ( i.e. $g =0$),  
all the  pairs occupy the lowest  energy levels forming a  
 Fermi sea. The pairing interaction promotes
the  pairs to  higher  energies  and eventually,
for large values of $N$, all the levels are pair correlated,
giving rise to   superconductivity \cite{BCS}.

\subsection*{Richardson's solution}

In order to describe Richardson's solution one
defines the hard-core boson operators

\begin{equation}
b_j = c_{j,-} c_{j,+}, \;\; b_j^\dagger= 
c^\dagger_{j,+} c^\dagger_{j,-} , \;\; N_j = b^\dagger_j b_j 
\label{2}
\end{equation}

\noindent which satisfy the commutation relations,

\begin{equation}
[ b_j, b_{j'}^\dagger ] = \delta_{j,j'} \; ( 1 - 2 N_j) 
\label{3}
\end{equation}

The Hamiltonian (\ref{1}) can then be written as

\begin{equation}
H_{BCS} = \sum_{j} 2 \varepsilon_j b^\dagger_j b_j - g \, 
\sum_{j,j'} \; b_j^\dagger b_{j^{\prime}} \; ,
\label{4}
\end{equation}

\noindent where we have set $d=1$ ( i.e. all the energies
are measured in units of $d$). 
Richardson showed that the eigenstates of this Hamiltonian 
with $N$ pairs have the (unnormalized) product form 
\cite{R1,R1bis,RS} ( for a direct proof of the results of this
subsection see reference \cite{ankara} )

\begin{eqnarray}
& |N \rangle_R = \prod_{\nu = 1}^N B_\nu |{\rm vac} \rangle, \;\;
B_\nu = \sum_{j=1}^\Omega \frac{1}{2 \varepsilon_j - e_\nu} 
\; b^\dagger_j&
\label{5} 
\end{eqnarray}

\noindent 
where the  parameters $e_\nu$ ($\nu = 1, \dots , N$) are, 
in general, complex solutions of the $N$ coupled algebraic
equations 

\begin{equation}
\frac{1}{g } + \sum_{\mu=1 ( \neq \nu)}^{N} \frac{2}{ e_\mu - e_\nu} 
= \sum_{j=1}^\Omega \frac{1}{2 \varepsilon_j - e_\nu} \; , 
\label{6}
\end{equation}

\noindent 
The energy of these states is given by the sum of the
auxiliary parameters $e_\nu$, i.e.

\begin{equation}
{\cal E} (N) = \sum_{\nu =   1}^{N} e_\nu
\label{7}
\end{equation}

\noindent 
The ground state  of $H_{BCS}$ is given by the solution
of eqs.(\ref{6}) which gives the lowest value of ${\cal E}(N)$. 
The (normalized) states (\ref{5}) can also be written  
as  \cite{R3}

\begin{equation}
|N \rangle_R = \frac{C}{ \sqrt{N!} } \sum_{j_1, \cdots , j_{N}}
\psi^R(j_1, \dots,  j_{N}) b^\dagger_{j_1} \cdots b^\dagger_{j_{N}}
|{\rm vac} \rangle 
\label{8}
\end{equation}

\noindent where 
the sum excludes double occupancy of pair states
and the wave function $\psi$ takes the form

\begin{equation}
\psi^R(j_1, \cdots, j_{N}) = \sum_{\cal P}
 \prod_{k=1}^{N} \frac{1}{ 2 \varepsilon_{j_k} 
- e_{{\cal P}k} }  
\label{9} 
\end{equation}

\noindent The  sum in (\ref{9})  runs  over all the 
permutations, ${\cal P}$, 
of $1, \cdots, N$. The constant  $C$ in (\ref{8})
guarantees the normalization of the state \cite{R3} ( i.e.
$_R\langle N|N\rangle_R =1$);  its expression will be given
in section IV.

A well known fact about the BCS Hamiltonian is that
it is  equivalent to that of a XY model with long
range couplings and a ``position dependent'' 
magnetic field proportional to $\varepsilon_j$.
To see this let us represent the hard-core boson operators
(\ref{2})  in terms of the Pauli 
matrices as follows,

\begin{equation}
b_j = \sigma_j^+, \; b_j^\dagger = \sigma_j^- ,\; 
N_j = \frac{1}{2} ( 1 - \sigma_z) 
\label{10}
\end{equation}

\noindent 
in which case  the Hamiltonian (\ref{4}) becomes

\begin{eqnarray}
& H_{BCS} = H_{XY} + \sum_j \varepsilon_j +
 g (\Omega/2 -N) & \label{11} \\
& H_{XY}= - \sum_j 2 \varepsilon_j t^0_j - \frac{g}{2}
( T^+ \; T^- + T^- \; T^+ ) & \nonumber 
\end{eqnarray}

\noindent where the matrices

\begin{eqnarray}
& T^a = \sum_{j=1}^\Omega  t^a_j \;\; (a = 0, +, -) & 
\label{12} \\
& t^0_j = \frac{1}{2} \sigma^z_j, \;\; t^+_j = \sigma^+_j ,
\;\; t^-_j = \sigma^-_j & \nonumber 
\end{eqnarray}

\noindent satisfy the $SU(2)$ algebra,

\begin{eqnarray}
& [ T^a, T^b] = f^{a b}_c T^b & \label{13} \\
& f^{+0}_+ = f^{0-}_- = -1, \; f^{+-}_0 = 2 & \nonumber 
\end{eqnarray}

\noindent whose   Casimir is given by

\begin{equation}
{\bf T} \cdot {\bf T} = T^0 T^0 +
\frac{1}{2} ( T^+ T^- + T^- T^+) 
\label{14}
\end{equation}

\subsection*{Integrability of the BCS Hamiltonian}

From the existence of an exact analytic solution of $H_{BCS}$, 
one may  expect that $H_{BCS}$  should be integrable.
Indeed  CRS   found 
the  integrals of motion \cite{CRS,note},

\begin{equation}
R_i = - t^0_i - g \; \sum_{j ( \neq i)}^\Omega
\frac{ {\bf t}_i \cdot {\bf t}_j }{ \varepsilon_i- \varepsilon_j} ,\;\;
(i=1, \dots, \Omega)
\label{15} 
\end{equation}

\noindent where the denominator does not blow up 
since we are assuming non degenerate energy levels. 
Integrability amounts to  the eqs.

\begin{eqnarray}
& [H_{BCS}, R_i] = 
 [R_i, R_j] = 0,  \;\;(i, j =1, \dots, \Omega) & \label{16} 
\end{eqnarray}

\noindent 
Denote the eigenvalue of $R_j$ acting on the state (\ref{8})
by  $\lambda_j$,  namely

\begin{equation}
R_j |N \rangle_R = \lambda_j |N \rangle_R
\label{17}
\end{equation}
 
\noindent 
CRS, seemingly unaware of Richardson's  
solution,  did not give  
an expression of $\lambda_j$ in their work.    
However,  they did show  that  $H_{XY}$ given in
eq.(\ref{11}) can be expressed in terms of the operators
$R_i$ as

\begin{equation}
H_{XY} = \sum_j 2 \varepsilon_j R_j + g ( \sum_j R_j )^2 
- \frac{3}{4} g \Omega 
\label{18}
\end{equation}

\noindent 
Hence,  combining (\ref{17}) and (\ref{18}) 
one can find the eigenvalues of $H_{XY}$

\begin{equation}
{\cal E}_{XY} = 
\sum_j 2 \varepsilon_j \lambda_j + g ( \sum_j \lambda_j )^2 
- \frac{3}{4} g \Omega 
\label{19}
\end{equation}

\noindent and in turn those of ${\cal E}_{BCS}$ by   
recourse to  eq.(\ref{11}).
We shall show in section IV  that  
 $\lambda_j$  has the  simple expression

\begin{equation}
\lambda_i = - \frac{1}{2} + 
g \left( \sum_{\nu=1}^N \frac{1}{ 2 \varepsilon_i - e_\nu}
- \frac{1}{4} \sum_{j=1 (\neq i)}^{\Omega}  \frac{1}{ 
\varepsilon_i - \varepsilon_j} \right) 
\label{20}
\end{equation}

\noindent 
One can  check  this result by
deriving  the energy  (\ref{7}) from eqs.  
(\ref{11}), (\ref{19}) and  (\ref{20}).

This ends the presentation of the exact solution of  the BCS
Hamiltonian. The existence of an underlying analytic
structure reminiscent to that of a CFT 
is apparent from eqs. (\ref{9}), (\ref{15})
and (\ref{20}). Indeed,  
the aforementioned 
equations contain factors of  the form
$1/(z - z')$ where $z$ and $z'$ stands for either $2 \varepsilon_j$
or $e_\nu$. Terms of this sort  arise quite naturally as
correlators ( i.e. $\langle A(z) B(z') \rangle = 1/(z-z')$)
of chiral primary fields $A(z)$ and $B(z)$ in diverse
CFT's. The problem  is to identify which CFT explains
all the features presented
so far in a unified manner. We shall argue that the solution of 
this problem  is given by 
the Wess-Zumino-Witten (WZW) model based on the affine Kac-Moody
group $SU(2)_k$ in the limit where the level $k$ goes
to $-2$. The proof of this result requires standard
tools of CFT and, more precisely,   the free field 
or Coulomb Gas (CG) representation of the WZW model.

\section*{III) Review of the 
free field representation of the
$SU(2)_k$-WZW model}

The material presented in this section is standard
in the CFT literature \cite{G}.
Nevertheless, we have  included it for the benefit
of readers that are not experts in CFT. 
This will allows us to highlight the main 
tools we shall use in later sections. 
We shall follow closely reference \cite{D}.

The WZW model is an interacting theory which nevertheless admits
a description in terms of free fields \cite{W}. 
The correlators and conformal
blocks can then be easily calculated as integrals of vacuum 
expectation values of 
vertex operators. This gives an integral representation of the conformal
blocks which satisfy automatically the Knizhnik-Zamolodchikov (KZ)
equations \cite{KZ}. In the case of the $SU(2)_k$-WZW model the free
fields are a $\beta-\gamma$ system with conformal weights
1 and 0,  respectively,  
and a boson field
$\varphi$ which satisfy the following  
operator product expansion (OPE) \cite{D,W},

\begin{eqnarray}
& \beta(z) \; \gamma(w) = - \gamma(z) \; \beta(w) = \frac{1}{z-w} &
\label{21} \\
& \varphi(z) \; \varphi(w) = - {\rm ln}(z-w) & \nonumber 
\end{eqnarray}

\noindent
The WZW currents $J^a(z) ( a=0, \pm)$ can be expressed  
in terms of these fields as ( hereafter, normal order of
operators will be  implicitely assumed)

\begin{eqnarray}
& J^+ = {\rm i} \; \beta & \nonumber \\
& J^0  = - \frac{ {\rm i}}{ 2 \alpha_0} 
\partial \varphi - \beta \; \gamma & \label{22} \\
& J^- = {\rm i} \;  [ \beta \gamma^2 +  \frac{ {\rm i}}{  \alpha_0}
\gamma \; \partial \varphi - k \partial \gamma ] & \nonumber 
\end{eqnarray}

\noindent which satisfy the OPE's

\begin{eqnarray}
& J^a(z) \; J^b(w) = \frac{k/2}{(z-w)^2} \; q^{ab}
+ \frac{1}{z-w} \; f^{ab}_c \; J^c(w) + {\rm reg.}\;{\rm terms} & 
\label{23}
\end{eqnarray}

\noindent where $f^{ab}_c$ are the $SU(2)$ structure
constants defined in eq.(\ref{13}),  and $q^{00}=1, \; q^{+-}=
q^{-+}= 2$. The level of the WZW model,  $k$, 
is related to the ``charge'' $\alpha_0$ by the eq.

\begin{equation}
k+2 = \frac{1}{2 \alpha_0^2 } 
\label{24}
\end{equation}

If $k$ is a positive integer,  the WZW model is a Rational
Conformal Field Theory (RCFT) with $k+1$ primary fields
labelled by the total spin, $j=0, 1/2, \dots, k/2$. 
In our CFT approach to BCS,  
we shall need to consider the limit
where $k \rightarrow -2$, which corresponds to taking 
$\alpha_0 \rightarrow \infty$. This is a singular limit
which takes us away from the rational WZW models. 
Actually, the 
case when $k$ is exactly $-2$ is mathematically 
interesting due to its relation  to the singular
hyperplanes in the representation theory of affine Kac-Moody
algebras \cite{KK,FF}. 
For non-positive integer values of $k$,  
we can still define the theory by the free
field representation given above. 

The Sugawara 
energy-momentum tensor $T_{Sug}$ of the WZW model is given 
by the sum of the energy momentum tensors of the
$\beta-\gamma$ system and the bosonic field $\varphi$

\begin{eqnarray}
& T_{Sug} = \beta \; \partial \gamma  
- \frac{1}{2} ( \partial \varphi )^2 
+ {\rm i} \alpha_0 \; \partial^2  \varphi & \label{25} 
\end{eqnarray}

The central extension $c$ of the Virasoro algebra
generated by the modes 
of $T_{Sug} ( = \sum_n L_n z^{-n-2} )$ is

\begin{equation}
c= 3 - 12 \alpha_0^2 = \frac{3 k}{k+2} \label{26}
\end{equation}

\noindent where the $\beta-\gamma$ system contributes
with 2 and the field $\varphi$ contributes with $1- 12 \alpha_0^2$.
In (\ref{26}) we have used the relation (\ref{24}),
which for integer $k$'s gives the well known value of the
Virasoro central charge of the $SU(2)_k$-WZW model. 
In the limit $(k+2) \rightarrow 0$ the central
extension $c$ diverges. In order to get  a meaningful theory 
one has  to scale the Virasoro operators as 
$\tilde{L}_n = lim_{k \rightarrow -2} (k+2)  L_n$. 
In that limit the Virasoro algebra  becomes

\begin{equation}
 [\tilde{L}_n , \tilde{L}_m ] = 0 
\label{27}
\end{equation}

\noindent which suggests some sort of integrability. In fact 
the commutativity of the Virasoro operators $\tilde{L}_n$ 
  has been used to study the representation theory
of the $SU(2)_{k=-2}$ Kac-Moody algebra for the 
proof of  the Kac-Kazhdan conjecture \cite{KK} concerning 
character formulas (see \cite{FF} for references).

The primary fields $\Phi^j_m(z)$  of the WZW model
are labelled by the total spin $j=0, 1/2, \dots$ and the third
component of the spin $m = j, \dots, -j$.
Their free field representation  is given by

\begin{eqnarray}
& \Phi^j_m(z)  = \gamma^{j-m}(z)\;  V_{\alpha_j}(z) & \label{28} \\
& V_{\alpha_j}(z) = e^{ {\rm i} \alpha_j \varphi(z)} 
,\;\; \alpha_j = - 2 \alpha_0 j & \nonumber
\end{eqnarray}

\noindent and have a conformal weight $\Delta_j$
given entirely by
that of the vertex operator $V_{\alpha_j}$, namely

\begin{equation}
\Delta_j = \frac{1}{2}  \alpha_j ( \alpha_j - 2 \alpha_0)
= \frac{j(j+1)}{k+2} 
\label{29}
\end{equation}

In the free field representation of a CFT,  every primary field
has a conjugate version besides its ``direct representant'', 
which is needed for the computation of correlators \cite{DF}. In the
case of the WZW model,  the conjugate of the primary field
with $m=j$ is

\begin{equation}
\tilde{\Phi}^j_j(z) = \beta^{s+2j}(z) \; V_{2 \alpha_0(s+j)}(z), 
\;\; s = -(k+1) \label{30}
\end{equation}

\noindent The corresponding equation for $m < j$ is much more complicated,
and it is a sum of terms where the difference between 
$\beta$ fields and $\gamma$ fields is given by $s+j+m$.

A  particular case of (\ref{30}) is when $j=0$, which corresponds
to the  conjugate field of the identity

\begin{equation}
\tilde{I}(z) = \beta^{s}(z) \; V_{2 \alpha_0 s}(z) \label{31}
\end{equation}

\indent 
A consequence of (\ref{31}) is that the expectation values 
of operators should satisfy the 
following charge neutrality conditions

\begin{equation}
N_\beta - N_\gamma = s, \;\; \sum_i \alpha_i = 2 \alpha_0 s,
\;\; s = -(k+1) 
\label{32}
\end{equation}

\noindent where $N_\beta$ and $N_\gamma$ is the number
of $\beta$ and $\gamma$ fields in the correlator, and
$\alpha_i$ are the charges of the vertex operators
made of the field $\varphi$.
Eq.(\ref{32}) means that there is a background charge
$- 2 \alpha_0 s$ in the boson sector and a charge $-s$
in the $\beta-\gamma$ sector, which need to be neutralize
for the correlator to be non-vanishing. The latter properties
can alternatively be attributed to the out vacuum which have
charges $-2 \alpha_0$ and $-s$ in the $\varphi$ and $\beta-\gamma$
sectors,  respectively. 

The remaining ingredient of the free field representation 
is provided by the so called screening 
charge

\begin{equation}
Q = \oint_C \; du S(u),\;\; S(u) = \beta(u) \; V_{2 \alpha_0}(u)
\label{33}
\end{equation}

\noindent whose basic property is that it commutes with the 
$SU(2)$ current algebra and the Virasoro operators. 
In eq.(\ref{33}) and below,  $du$ is meant to contain
the factor $1/(2 \pi {\rm i})$ to takes care of the factor
$2 \pi {\rm i}$ that comes out  in the residue formula. 
Using the vertex representations (\ref{28}) and (\ref{30})
of the primary fields,  together with the screening charge
(\ref{33}),  one can compute the conformal blocks of the WZW model.
Conformal blocks are the chiral building blocks of 
correlators. The latter are obtained by combining the holomorphic
and the anti-holomorphic conformal blocks and imposing monodromy
invariance. In the WZW model a conformal block 
$\psi^{WZW}(z_1, \dots, z_{\Omega+1})$
involving 
$\Omega+1$ primary fields $\{(j_k,m_k)\}_{k=1}^{\Omega+1}$,
inserted at the positions $\{ z_k\}^{\Omega+1}_{k=1}$, 
can  be associated 
with  the $SU(2)$ tensor product decomposition

\begin{eqnarray}
& j_1 \otimes  \dots \otimes j_\Omega \rightarrow j_{\Omega+1} & \label{34} \\
& \sum_{k=1}^{\Omega} \; m_k = m_{\Omega+1} & \nonumber
\end{eqnarray}

\noindent \noindent where  $(j_{\Omega+1},m_{\Omega+1})$ appears
as an outgoing state. 
The free field expression of the conformal blocks is given by

\begin{eqnarray}
& \psi^{WZW} 
(z_1, \dots, z_{\Omega+1})
& \nonumber \\ 
& = \langle \Phi_{m_1}^{j_1}(z_1) \dots 
\Phi_{m_\Omega}^{j_\Omega}(z_\Omega) \; 
\tilde{\Phi}_{m_{\Omega+1}}^{j_{\Omega+1}}(z_{\Omega+1}) 
& \label{35} \\
& \times \oint_{C_1} du_1 S(u_1)
\dots \oint_{C_N} du_N S(u_N) \rangle & \nonumber 
\end{eqnarray}

\noindent where  $\tilde{\Phi}_{m_{\Omega+1}}^{j_{\Omega+1}}$ 
is the conjugate of the outgoing state $(j_{\Omega+1}, m_{\Omega+1})$,
and the screening charges are 
integrated 
along the contours $C_1, \dots, C_N$. The charge neutrality conditions
(\ref{32}) applied to  (\ref{35}) yield

\begin{equation}
N = \sum_{k=1}^{\Omega} j_k - j_{\Omega+1} ,\;\; 
m_{\Omega+1} = \sum_{k=1}^\Omega m_k 
\label{36}
\end{equation}

\noindent which agree with the Clebsch-Gordan decomposition
(\ref{34}). 
The case when $N=0$ corresponds to the maximal allowed
value of $j_{\Omega+1}= \sum_{k=1}^\Omega j_k$. 
On the other hand,  if $\Omega$ is
even,  the minimal  value of $j_{\Omega+1}$ is zero,  
which requires $N= \sum_{k=1}^{\Omega} j_k$ screening charges.
Hence,  the different choices of the screening charges 
and contours give rise to all possible  conformal blocks. 
In this manner the free field representation provides  
integral solutions of the KZ equations satisfied by the 
conformal blocks (\ref{35}). The KZ eqs are \cite{KZ}

\begin{equation}
\left( \kappa \frac{\partial}{\partial z_i} 
- \sum^{\Omega +1}_{ j=1 ( \neq i)  }
\frac{ {\bf t}_i \cdot {\bf t}_j }{ z_i - z_j} 
\right) \psi^{WZW}(z_1, \dots, z_{\Omega+1} )
\label{37}
\end{equation}

\noindent where 
$\kappa = (k +2)/2 $ and ${\bf t}_i$ are 
the $SU(2)$ matrices in the $j_i$ representation 
acting at the $i^{\rm th}$ site.

\section*{IV) CFT representation of the exact solution of BCS}

Our first aim is to  obtain Richardson's
wave function (\ref{9}) using the free field representation
of the WZW model.

\subsection*{Richardson's wave function}

The starting point is
the  speudospin version of the BCS
model introduced in section II, 
according to which  an empty  energy level 
$\varepsilon_i ( i=1, \dots, \Omega)$
has spin $m_i= 1/2$ while and occupied level has spin $m_i = -1/2$. 
This suggests to rewrite 
Richardson's wave function as

\begin{equation}
\psi_{m_1, \cdots, m_{\Omega}}^R ( z_1, \dots, z_\Omega;e_1, \dots, e_N)
 = \sum_{\cal P}
 \prod_{k=1}^{N} \frac{1}{ z_{l_k} 
- e_{{\cal P}k} }   
\label{38} 
\end{equation}

\noindent where $z_i = 2 \varepsilon_i$ and 
$l_k$ is defined by the condition $m_{l_k} = -1/2$. 
The $SU(2)$ quantum numbers $m_i$ of  (\ref{38})  
satisfy

\begin{equation}
\sum_{i=1}^\Omega m_i = \frac{\Omega}{2} - N
\label{39}
\end{equation}

Let us compare now the conformal block (\ref{35})
and the wave function (\ref{38}). If we take $j_k=1/2
(k=1, \dots, \Omega)$ in (\ref{35}) and use 
(\ref{39}), 
we are lead to the identifications

\begin{equation}
j_{\Omega +1} = m_{\Omega+1} = \frac{\Omega}{2} - N 
\label{40}
\end{equation}

\noindent which requires $N \leq \Omega/2$. 
Hence from a formal point of view, 
we can regard Richardson's wave function as 
a conformal block involving  
$\Omega$ primary fields $\Phi^{1/2}_{m_j}(z_j)$, 
located at  the positions $z_j = 2 \varepsilon_j$,
and a primary field 
$\tilde{\Phi}^{j_{\Omega +1}}_{m_{\Omega +1}}(z_{\Omega +1})$  
whose position we shall place
at $\infty$. 
The last ingredient we need, in order 
to reproduce Richardson's
wave function using CFT tools, is to find the role
played by the BCS coupling constant $g$. 
We shall see below that $g$ is associated to the
operator

\begin{equation}
V_{g}= {\rm exp} (  {- \frac{ {\rm i} \alpha_0}{ g} \oint_{C_g} 
dz \; z \partial \varphi(z)} ) 
\label{41}
\end{equation}

\noindent 
Our claim is that  
Richardson's wave function  (\ref{38})
is given, up to a proportionality factor,
by the limit

\begin{equation}
\psi^{R}_{\bf m}({\bf z},{\bf e}) \propto
\lim_{\alpha_0  \rightarrow \infty} 
\psi^{CG}_{\bf m}({\bf z}) 
\label{42}
\end{equation}

\noindent where the Coulomb Gas wave function 
$\psi^{CG}$ is given by the following 
expectation value

\begin{eqnarray}
& \psi^{CG}_{{\bf m} }
({\bf z}) & \nonumber \\
& = \langle V_g  \Phi_{m_1}^{\frac{1}{2}}(z_1) \dots 
\Phi^{\frac{1}{2}}_{m_\Omega}(z_\Omega) & \label{43} \\ 
&\times  \tilde{\Phi}_{j_{\Omega+1}}^{j_{\Omega+1}}(\infty) 
\; \oint_{C_1} du_1 S(u_1)
\dots \oint_{C_N} du_N S(u_N) \rangle & \nonumber 
\end{eqnarray}

\noindent 
Except for the presence of the operator $V_g$, 
eq.(\ref{43}) coincides with the conformal block 
(\ref{35}). Let us now prove eq.(\ref{42}). 
First of all,  using the free field representation
of the primary fields,  one can write (\ref{43}) as

\begin{eqnarray}
& \psi^{CG}_{{\bf m} }
({\bf z}) = \oint_{C_1} du_1 \dots \oint_{C_N} du_N 
 \psi^{\varphi}({\bf z}, {\bf u}) 
\;  \psi^{\beta \gamma}_{{\bf m}} ({\bf z}, {\bf u}) &
\label{44}
\end{eqnarray}

\noindent where

\begin{eqnarray}
& \psi^\varphi   
= \langle V_g  \prod_{i=1}^\Omega 
V_{- \alpha_0}(z_i)
\prod_{\nu =1}^N V_{2 \alpha_0}(u_\nu)
V_{2 \alpha_0 (s + j_{\Omega+1}) }(\infty) \rangle &      
\label{45} 
\end{eqnarray}

\[
\psi^{\beta \gamma}_{\bf m} 
=  \langle \prod_{i=1}^\Omega \gamma^{\frac{1}{2}-m_i}(z_i) 
\prod_{\nu=1}^N \beta(u_\nu) 
\beta^{s + 2 j_{\Omega+1}}(\infty)
\gamma^{2 j_{\Omega+1}}(\infty) \rangle 
\]

\noindent
Using eqs.(\ref{21}) and the Wick theorem, 
one can show that 
$\psi^{\beta \gamma}$
is,  up to a sign, equal  
to the Richardson's wave function, namely

\begin{eqnarray}
\psi^{R}_{\bf m}({\bf z}, {\bf e}) = (-1)^N
\psi^{\beta \gamma}_{\bf m}({\bf z}, {\bf e}) &
\label{46}
\end{eqnarray}

In this equation  we have choosen the positions of the
screening operators  $u_i$ equal to the Richardson's 
parameters $e_i$. However,  in eq.(\ref{44})
one must integrate over the $u's$. 
The job of the limit 
$\alpha_0 \rightarrow \infty$ is to set $u_i = e_i$. 
Let us see how this happens.
First of all  the contribution of the vertex operators
can be computed using the formula

\begin{eqnarray}
& \langle V_g  V_{\alpha_1}(w_1) \dots V_{\alpha_M}(w_M)\rangle
& \label{47} \\
& = \prod_{i<j} \; (w_i - w_j)^{\alpha_i \alpha_j}
\; \prod_j  e^{ - \alpha_0 \alpha_j w_j / g}
& \nonumber 
\end{eqnarray}

\noindent where the contour $C_g$ encircles all the
coordinates $w_j$. The factor $\psi^\varphi$ then becomes

\begin{eqnarray}
& \psi^{\phi}({\bf z}, {\bf u}) 
= \prod_j  e^{  \alpha_0^2 z_j / g}
\prod_\nu e^{ -2 \alpha^2_0 u_\nu/g}  & \label{48} \\
& \times 
\prod_{i<j} (z_i - z_j)^{\alpha^2_0}\;
\prod_{\nu < \mu} (u_\nu - u_\mu)^{4 \alpha_0^2}
\prod_{i, \nu}  (z_i - u_\nu)^{-2 \alpha_0^2}
& \nonumber 
\end{eqnarray}

\noindent 
In the limit  $\alpha_0 \rightarrow \infty$, the integral
(\ref{44}) can be computed using the saddle point method. 
Indeed,  writting (\ref{48}) as

\begin{eqnarray}
& {\psi}^{\varphi}({\bf z}, {\bf u})=  
e^{ - \alpha_0^2 \;  U({\bf z}, {\bf u}) } & \label{49}
\end{eqnarray}

\noindent where 

\begin{eqnarray}
& U = - \sum_{i < j}^\Omega {\rm ln}(z_i - z_j) - 4 
\sum_{\nu < \mu}^N {\rm ln}(u_\nu - u_\mu) & \label{50}\\
& + 2 \sum_{i=1}^\Omega \sum_{\nu=1}^N {\rm ln}(z_i - u_\nu) 
+ \frac{1}{g} ( - \sum_{i=1}^\Omega z_i + 2 \sum_{\nu=1}^N
u_\nu ) & \nonumber 
\end{eqnarray}

\noindent 
the stationary solutions  of $U$ are given 
by the solutions of Richardson's eq.(\ref{6}), namely

\begin{equation}
0= \left( \frac{\partial U}{ \partial u_\nu} \right)_{u_\mu = e_\mu} 
\rightarrow  
\frac{1}{g } + \sum_{\mu=1 ( \neq \nu)}^{N} \frac{2}{ e_\mu - e_\nu} 
= \sum_{j=1}^\Omega \frac{1}{z_j - e_\nu} \; , 
\label{51}
\end{equation}

\noindent 
Under these conditions the saddle point value of the
integral (\ref{44}) is

\begin{eqnarray}
& \psi^{CG}_{\bf m}( {\bf z}) \sim 
\frac{1}{ (2 \pi^{1/2} \alpha_0 )^N ({\rm det} {\bf A})^{1/2} } \; 
e^{ - \alpha^2_0 U({\bf z}, {\bf e}) } \; 
\psi^{\beta \gamma}_{\bf m}({\bf z}, {\bf e}) &
\label{52}
\end{eqnarray}

\noindent where ${\bf A}$ is the $N \times N$ hessian matrix 
defined as

\begin{eqnarray}
& {\bf A}_{\mu, \nu} = - \frac{1}{2} \frac{ \partial U}{ 
\partial u_\mu \; \partial u_\nu}|_{{\bf u} = {\bf e}} &
\label{53} \\
& A_{\nu, \nu} = \sum_i \frac{1}{ (z_i - e_\nu)^2} 
-  \sum'_\nu \frac{2}{ (e_\nu - e_\mu)^2}, \; 
A_{\mu, \nu} =  \frac{2}{ (e_\nu - e_\mu)^2} & \nonumber 
\end{eqnarray}

Eqs.(\ref{46}), (\ref{51})  and (\ref{52}) constitute  the proof of
(\ref{42}). Moreover the factor 
$ ({\rm det} {\bf A})^{-1/2}$ in (\ref{52}) 
turns out to coincide
with the normalization constant $C$ appearing 
in the normalized state (\ref{8}) \cite{R3}, namely

\begin{equation}
C = 1/  ({\rm det} {\bf A})^{1/2}
\label{54}
\end{equation}

Let us make some comments about the results 
presented so far.

\begin{itemize}

\item  Richardson has observed that eqs.(\ref{6})
could be derived as the stationary configurations for
a set of electrostatic charges in 2D \cite{R7}. The
electrostatic  potential of these charges is given
by $\alpha_0^2 U$ or rather 
 by the sum $\alpha_0^2(U + U^*)$.
Our  CFT derivation of the exact  BCS  solution  shows 
that the value of the  charges are $- \alpha_0$ for the
energy levels,  $2 \alpha_0$ for the screening ones, 
and  a  charge $2 \alpha_0
(s + j_{\Omega +1})$ placed at infinity. The sum of all these 
charges neutralizes the background charge $- 2 \alpha_0 s$.
The stationary conditions are set by the limit 
$\alpha_0 \rightarrow \infty$, which is equivalent
to the limit $k \rightarrow -2$.

\item The operator $V_g$, which is needed in order
to get the term $1/g$ in the Richardson's equations,
breaks conformal invariance in an explicit manner. 
This is why $\psi^{CG}$ is not, strictly speaking,
a conformal block of the WZW model. In the spirit
of Perturbed Conformal Field Theory  
 \cite{Z} we could   say that 
the  WZW model has been perturbed by the chiral operator
$ - \frac{{\rm i} \alpha_0}{g} \oint dz z \partial \varphi$
which is equal to $- \alpha_0 a_{-1} /g$,  where 
$a_{-1}$ is the $n=-1$ mode of the field $\partial \varphi$ 
 (${\rm i} \partial \varphi(z) = \sum_n a_n z^{-n-1}$).

\item The perturbative renormalization group (RG) analysis 
of the BCS model yields that the coupling constant
$g >0 $ flows to large values in the infrared regime, 
leading to the superconducting
instability of the Fermi sea \cite{Shankar}.
In our approach  
the fixed point of this RG flow is described 
by  the $SU(2)_{k=-2}$-WZW model. The $1/g$ perturbation
takes us away from this fixed point,  breaking
conformal invariance.  In that sense,  we are dealing with
the strong coupling version of BCS, which is valid
for all values of the BCS coupling constant $g$,  since there
are not phase transitions as we go from  weak to  strong coupling.

\item So far,   we have assumed for simplicity 
that the pair degeneracy of the energy levels $\varepsilon_j$ 
is unity, which corresponds in the WZW model to primary fields
with spin $j_k= 1/2$. The construction can be straighforwardly
generalized  to levels with higher 
degeneracy,  $d_k=1,2, \dots$, in which case 
the associated primary fields have spin $j_k = d_k/2$. 
In this case the Richardson equations   read  \cite{R6,R7},

\begin{equation}
\frac{1}{g } + \sum_{\mu=1 ( \neq \nu)}^{N} \frac{2}{ e_\mu - e_\nu} 
= \sum_{j=1}^\Omega \frac{d_j}{z_j - e_\nu} \; , 
\label{51bis}
\end{equation}

\noindent and they  can be derived using the free field 
representation explained above 
in terms of  the fields $\Phi^{j_k}_{m_k}(z_k)$ with
$j_k = d_k/2$.

\end{itemize}

\subsection*{The integrability of BCS from CFT}

At this stage it is clear that  eqs. (\ref{15})
and (\ref{17}) must be  related to the KZ eq. (\ref{37}).
After all, in the limit $\alpha_0 \rightarrow \infty$
the CG  wave function (\ref{43}) coincides with the WZW
wave function. Let us show this in detail. The KZ eq.(\ref{37})
with $z_{\Omega+1} = \infty$ and the definition
(\ref{15}) can be written as

\begin{equation}
\left( \kappa \frac{\partial}{\partial z_i} 
+ \frac{1}{2 g} \; R_i + \frac{1}{2 g} \; t^0_i  
\right) \psi^{WZW} = 0 
\label{55}
\end{equation}

\noindent 
Let us define  a new wave function $\psi$

\begin{equation}
\psi^{WZW} = e^{ H_{XY} / 2 g \kappa} \; \psi 
\label{56}
\end{equation}

\noindent Using the commutativity of  $R_i$ and  $H_{XY}$
and the fact that

\begin{equation}
 \frac{\partial}{ \partial z_i} H_{XY} = - t^0_i 
\label{57}
\end{equation}

\noindent  eq.(\ref{55}) becomes

\begin{equation}
\left( \frac{1}{2 g} \, R_i + \kappa  \frac{\partial}{ \partial z_i} 
\right) \psi = 0 
\label{58}
\end{equation}

This equation is completely equivalent to the KZ eq.
Imposing that $\psi$ diagonalizes $R_i$
with eigenvalue $\lambda_i$,  $\psi$  becomes
proportional to the Richardson solution,  or rather to 
 $\psi^{CG}$. Hence   
eq.(\ref{49}) yields  the following 
 expression for $\lambda_i$

\begin{equation}
\lambda_i = \frac{g}{2} \frac{ \partial U}{ \partial z_i}
\label{59}
\end{equation}

\noindent which in turn leads to  the formula
(\ref{20}). 
In other words, $\psi^{WZW}$, $\psi^{CG}$ and $\psi^R$
agree up to overall factors in the limit when $\alpha_0 
\rightarrow \infty $. It is interesting to observe that the
factor relating the WZW and  Richardson
wave functions  has indeed a conformal structure reflected 
solely in the  presence of logarithms, namely

\begin{eqnarray}
& \psi^{WZW} = e^{ \alpha^2_0 \left( \frac{2}{g} E_{XY} -U \right) }
\; \psi^R,  & 
\label{60} \\
&  \frac{2 E_{XY}}{g}  -U  = \sum_{i < j} {\rm ln} (z_i - z_j) &
\nonumber \\
& + 4 \sum_{\nu <  \mu} {\rm ln} (u_\nu - u_\mu) 
- 2 \sum_{i,\nu} {\rm ln}(z_i - u_\nu) & \nonumber 
\end{eqnarray}

The Coulomb energy $U$  contains in a disguised manner both 
the Richardson's eqs. and the constants of motion $\lambda_i$.

\section*{V) The BCS/CFT correspondence}

In the previous section we have shown the 
closed relationship between the
BCS and the CFT  wave functions. In this section we want
to investigate this relation at the second quantized
level. 
In a Quantum Field Theory 
a generic N-body wave function $\Psi(x_1, \dots, x_N)$
is usually constructed from the overlap of a $N-$body 
state $|\Psi\rangle$ with the eigenstates 
$|x_1, \dots, x_N\rangle$ created by the action
of the field operator $\hat{\psi}(x)$ acting 
on the Fock vacuum $|0 \rangle $, namely

\begin{eqnarray}
& \Psi(x_1, \dots, x_N) = \langle \Psi| x_1, \dots, x_N \rangle &
\label{61} \\
& |x_1, \dots, x_N \rangle = \hat{\psi}(x_1) \dots 
\hat{\psi}(x_N) | 0 \rangle & \nonumber
\end{eqnarray}

An interesting example of this formalism 
is  provided by  the CFT interpretation of the Laughlin wave
function \cite{L} of the Fractional Quantum Hall effect (FQHE), 
first proposed by Fubini \cite{F}. In that case 
$\hat{\psi}(x)$ is a vertex operator of a single
boson governed by a $c=1$ CFT. 
One can also make use of an  array  of bosons.

In the spirit of the FQHE/CFT correspondence,  
one can  interpret the Coulomb Gas wave function
(\ref{43}) as

\begin{equation}
\psi^{CG}_{\bf m}({\bf z}) = 
\langle \psi^{CG} | z_1, m_1,  \dots, z_\Omega, m_\Omega \rangle
\label{62}
\end{equation}

\noindent where

\begin{eqnarray}
& |z_1,m_1,  \dots, z_\Omega,m_\Omega \rangle   = 
\prod_{i=1}^\Omega  \Phi_{m_i}^{\frac{1}{2}}(z_i) | 0 \rangle& 
\label{63} \\
& \langle \psi^{CG} | = \langle \alpha_0 (\Omega - 2N) | V_g \;
\; \prod_{\nu=1}^N  \oint_{C_\nu} du_\nu S(u_\nu)  & \nonumber 
\end{eqnarray}

\noindent 
The out vacuum  $\langle \alpha_0 (\Omega - 2N) |$
has been defined as the action of the operator
$\tilde{\Phi}^{j_{\Omega +1}}_{m_{\Omega+1}}$ on the
out vacuum of the WZW model, and has a charge $\alpha_0 (\Omega - 2N)$
in the $\varphi$ sector and no charge in the $\beta-\gamma$ sector.
The different states $\langle \psi^{CG}|$ correspond
to different choices of the integration contours $C_\nu$
of the screening operators $S(u_\nu)$, which in  CFT
yield  different conformal blocks. For example,  the
ground state is obtained by choosing the 
$\nu^{\rm th}$-contour ($\nu = 1, \dots, N$)
to run from $z_\nu$ to infinity, where we assume that 
the energy levels are ordered in increasing order, i.e.
$z_1 < z_2 < \dots < z_\Omega$. The excited states
correspond to other contour choices. The total number
of eigenstates of $H_{BCS}$ with $N$ pairs 
and $\Omega$ levels, which is given by the combinatorial
number $\left( \begin{array}{c} \Omega \\ N \end{array} \right)$,
coincides  with the total number of contour choices.

Eqs.(\ref{62}) and (\ref{63}) provide the basic correspondences between 
the exact solution of  BCS  and CFT, which can be extended 
to other instances. We collect them in table  1 and make
some comments below.

\begin{center}
\begin{tabular}{|c|c|} 
\hline
\hline
BCS & CFT \\
\hline
\hline
Pair energy level & WZW Primary field \\
\hline
Pair degeneracy ($d_k$) & Total spin ($ j_k=d_k/2$) \\ 
\hline
Eigenstates of $H_{BCS}$ & Conformal blocks \\ 
\hline
Richardson's eqs & Saddle point conditions \\
\hline
Integrability (CRS) & KZ eqs. \\  
\hline
$g= \infty$  ( $g$ finite) & WZW ( Perturbed WZW) \\
\hline
Cooper pair operator & Screening operator \\
\hline
Phase stiffness & $\propto \alpha_0^2$ \\ 
\hline
empty, occupied  level & spin up, down \\
\hline
$\prod_{k=1}^N b^\dagger_{j_k} |{\rm vac} \rangle $
& $ \prod_{i=1}^\Omega \Phi^{1/2}_{m_i}(z_i) | 0 \rangle $ \\
\hline 
$_R\langle N|$ &
$\langle \psi^{CG}| $ \\
\hline  
$\psi^R$ & $\psi^{\beta \gamma}$ \\
\hline
$b_j, b_j^\dagger, \frac{1}{2} -N_j$ & 
$ \oint_{z_j} dz J^+(z), J^-(z), J^0(z) $ \\
\hline  
C & $1/\sqrt{ {\rm det} {\bf A}}$ \\
\hline
$ \lambda_i$  & 
$ \frac{ g}{2} 
\frac{ \partial U_C}{ \partial z_i}$ \\
\hline 
$R_i$ & $- g \tilde{L}_{-1}^{(i)}$ \\
\hline
$H_{XY}$ & $-g \tilde{L}_0 $ \\
\hline   
\end{tabular}
\end{center}
\begin{center}
Table 1. The BCS/CFT correspondence 
\end{center}

\begin{itemize}

\item In analogy to eqs.(\ref{63}) 
one could try to define the bra state$
 \langle z_1,m_1,  \dots, z_\Omega,m_\Omega | $
using the conjugate operators
$\tilde{\Phi}^{\frac{1}{2}}_{m}(z)$. However,  this
state would have a large charge which would be
difficult to compensate for. Similarly,  
the ket state
$| \psi^{CG} \rangle$ is not easy 
to construct since the screening operator
$S(z)$ does not have a conjugate version. 
A possibility would be to use 
the  second screening operator,
given by 
$\tilde{S}(z) = \beta^{s-1} {\rm exp}(- {\rm i} 
\varphi(z)/\alpha_0)$ \cite{D}.
This operator appears in the 
construction of  fractional spin representations
which are related upon a certain reduction to the
minimal models. In the limit $\alpha_0 \rightarrow \infty$
we see that $\tilde{S}(z)$ converges to the identity. 
Further work is needed to clarify this issue.

\item In section III we defined the Virasoro
operators $\tilde{L}_n$ 
 as the singular limit  $\lim_{k\rightarrow -2} (k+2) L_n$.
From eq. (\ref{58}) we  see that $R_i$ could be
identified with the action of $- g \tilde{L}_{-1}^{(i)}$ on
$\psi^{CG}$, where  $  \tilde{L}_{-1}^{(i)} = \partial/\partial z_i$.
Similarly,  eq.(\ref{18}) implies  that 
$H_{XY}$ could be identified, up to constants, with
$- g (k+2) \sum_j z_j \partial/ \partial z_j$
and thus with 
$ - g \tilde{L}_0$. Upon these identifications,  the
integrability of the BCS model, given by eqs. (\ref{16}),  
becomes equivalent to 
the commutativity (\ref{27}) 
of the Virasoro operators
$\tilde{L}_0$ and $\tilde{L}_{-1}^{(i)}$.   
We  may expect the existence
of another CRS-like integrals of motion associated
to the Virasoro operators $\tilde{L}_n (n \leq -2)$.

\item The Richardson's  state $_R \langle N|$ 
corresponds, in the CFT formulation,  to the 
state $\langle \psi^{CG}|$, which is the product
of $N$ screening operators acting on the 
out vacuum  $\langle \alpha_0 (\Omega - 2N) |$. 
This correspondence suggests that in the grand
canonical (g.c.) ensemble ( where the number of pairs
is not fixed )  the corresponding state should 
be given by

\begin{equation}
\langle \psi^{CG}_{g.c.}| = \langle 0 | \; {\rm exp}( { \oint dz S(z)} ) 
\label{gc}
\end{equation}

\noindent 
( Note that  we have assumed the half filled condition 
$\Omega = 2 N$).  This state is similar to the 
BCS state given by \cite{BCS} 

\begin{equation}
\langle BCS| = \langle {\rm vac}| \; {\rm exp} ({ \sum_k g_k b_k}) 
\label{bcs}
\end{equation}

\noindent where $g_k$ is the  ratio $v_k/u_k$ 
of the BCS variational parameters. In this sense, 
the screening charge  $ \oint dz S(z)$ is the CFT version
of the Cooper pair operator  $ \sum_k g_k b_k $. 
In CFT it has been argued \cite{DF} 
that the screening operators can be exponentiated 
into the action and that their number is fixed 
upon imposing 
the charge neutrality conditions on correlators. 
This CFT  exponentiation corresponds 
to working in the grand canonical emsemble in BCS.

\item In the previous item we argued that $S(z) = \beta(z)
{\rm exp}(2 \alpha_0  {\rm i} \varphi(z))$ is the CFT 
analogue of $g_k b_k$. On the other hand,  $b_k$ corresponds
to the contour integration of $J^+(z) = {\rm i} \beta(z)$
around  the pair energy $z_k = 2 \varepsilon_k$.
Hence it is natural to associate the BCS variational parameter
$g_k$ with ${\rm exp}(2 \alpha_0  {\rm i} \varphi(z))$. 
This  means that $2 \alpha_0 \varphi(z)$
can be associated to the phase of 
the superconducting  order parameter. 
Shifting $\varphi(z)$ by a constant leads to an overall 
phase shift
of the BCS order parameter. This correspondence yields 
an insight about the physical meaning of $\alpha_0^2$, which
seems to be related to the phase stiffness or the superfluid
density $n_s$. Indeed,  if we identify the phase of the superconducting
order parameter $\theta$ with  $\alpha_0 \varphi$ then 
the Lagrangian of $\varphi$ becomes  that of a continuum 
$XY$  model 
for $\theta$, with $\alpha_0^2$ playing the role of the superfluid
density. Actually $\alpha_0^2$ appears in the denominator
of the Lagrangian while $n_s$ appears in the numerator. However
recall that we are working in the energy space so things are inverted.
The identification of $\theta$ with  $\alpha_0 \varphi$ is also
consistent with the fact that both variables are defined
modulo $2 \pi$. The limit $\alpha_0 \rightarrow \infty$
therefore corresponds to the limit of very large
phase stiffness which in fact leads to the standard
BCS theory, where the phase of the superconducting order
parameter is rigid and plays no role in fixing
the critical temperature or other observables \cite{phase}.
As was shown by Richardson \cite{R7} eqs.(\ref{6}) reduce
in the bulk limit $N \rightarrow \infty$ to the BCS gap equation
and hence  the state (\ref{8}) becomes  the fixed $N$ projection 
of the mean field g.c. BCS state. Finite values
of $\alpha_0$ should  lead  to non mean field
theories with   the phase $\theta$  playing a dynamical
role. It is rather intriguing that models of this sort have
already been proposed by several authors for  
an  explanation of  high-$T_c$ superconductivity \cite{phase}.

\end{itemize}

\section*{VI) Comparison between BCS and the FQHE}

In sections V and VI we noticed some analogies between
the CFT approaches to  BCS and the FQHE. Let us consider
them in some more detail. A common feature is the 
Coulomb Gas treatment. 
In the FQHE the CG is associated to  the Laughlin
wave function of $N_e$ electrons at filling
factor $\nu = 1/m$, \cite{L}

\begin{equation}
\psi_L(w_1, \dots , w_{N_e}) = \prod_{i<j}^{N_e} (w_i - w_j)^m 
e^{ - \frac{1}{4} \sum_\ell |w_\ell|^2 } 
\label{64}
\end{equation}

The norm of (\ref{64}) can be seen as a classical
probability distribution $e^{- \beta U_L}$  of a 
two-dimensional  one-component 
plasma at ficticious temperature
$\beta = 1/m$ and  potential energy $U_L$ where  \cite{L},

\begin{equation}
U_L = - 2 m^2 \sum_{i<j} {\rm ln} |w_i - w_j| + \frac{m}{2}
\sum_{\ell}^{N_e} |w_\ell|^2
\label{65}
\end{equation}

\noindent The particles with charge $m$ repell each other
with a logarithmic interaction,  and they are attracted 
to the origin by an uniform neutralizing background charge
with density $\rho= 1/(2 \pi \ell^2_B)$, where $\ell_B$ is the
magnetic length, which has been set equal to one in 
(\ref{64}) and (\ref{65}). 
For small values of $m (=3,5, \dots)$ the electrons form a liquid
with uniform density $\rho_e = 1/(2 \pi m \ell_B^2)$ which
neutralizes the background charge. However,  for large values
of $m$, Quantum Montecarlo studies have shown that
the Laughlin liquid becomes a solid ( i.e. a Wigner crystal)
where the positions of the charges are localized.

The comparison between  the wave function $\psi^{\varphi}$ given in eq.
(\ref{48}) and  the Laughlin wave function $\psi_L$ suggests
a formal   identification of  the electron positions $w_j$  with the 
screening positions $u_j$ rather 
than with the pair energies levels $z_j$. 
The reason is that both
the $u's $ and the $w's$ are subject to integration, while
the $z's$ are held fixed. Following this analogy,  
we may stablish the relations

\begin{equation}
m = 4 \alpha_0^2, \;\; {\rm or} \;\; \nu = \frac{k+2}{2} 
\label{66}
\end{equation}

\noindent according to which the freezing  of the 
screening charges in the limit $\alpha_0  \rightarrow \infty$  
would parallel  the Wigner crytal structure of the FQHE
when $m$ is large. 
On the contrary, for finite values of  $\alpha_0$ 
the screening charges, which are essentially Cooper pairs,  
would delocalize becoming
a sort of  liquid. The discussion at the end
of the previous section suggests  that this liquid
should arise from the 
fluctuations of  the  phase of the superconducting
order parameter.

Besides these analogies,  
the Laughlin and BCS Coulomb Gas models differ 
in the nature of the background charge. In the Laughlin
case the charge is two-dimensional,  while in the BCS
case the linear terms appearing in $U$ ( see eq. (\ref{50})) 
can be attributed to a linear uniform density $\rho_\ell
\propto \alpha_0/g$ placed at infinity. The latter density creates
a uniform electric field $\partial_x  (U + U^*)$ along the $x= (z + z^*)/2$
axis.  Another difference is that the BCS theory is not really  conformal
invariant for finite values of $g$ 
while the Laughlin state has gapless edge excitations
described by CFT \cite{Wen}.

\section*{VII) Conclusions and Prospects}

In this paper we have established a closed relationship
between the exact solution of the BCS Hamiltonian 
and the Coulomb Gas version of the 
$SU(2)_k$-WZW model in the singular 
limit when $k \rightarrow -2$. The Richardson's wave
function comes from the $\beta-\gamma $ chiral correlators, 
while the Richardson's eqs. and the normalization factor
of the state arises from the saddle point
evaluation of the chiral boson correlators. 
The BCS coupling constant $g$ enters the construction 
as a perturbation of the WZW model and 
breaks  conformal invariance. The integrability
of the BCS model is related to that of the WZW model
through the KZ equations, which has lead us to 
an expression of 
the integrals of motion of the BCS model 
found by Cambiaggio, Rivas and Saraceno.  
We have  proposed a  BCS/CFT correspondence which,
in many respects,   
parallels the 
CFT interpretation of the Fractional Quantum Hall effect. 
We have conjectured that the
singular limit $\alpha_0^2 \rightarrow \infty$ 
amounts  in physical terms  
to the limit of very strong phase stiffness, 
which leads to the  mean field BCS theory. 
Finite $\alpha_0$ generalizations of the
BCS model may  correspond  to non mean field
theories where   the phase of the superconducting
order parameter should have    a dynamical role 
as in some models of high-$T_c$ 
superconductivity \cite{phase}.

Besides giving new insights into the exact solution of the BCS
model, the CFT approach may also help in solving some 
problems as the computation of observables with 
the Richardson's exact solution.
The finite temperature BCS model is also an interesting problem
which one may try to  address with CFT tools. 

Finally,  the BCS/CFT approach can be straighforwardly 
generalized  to any WZW model based on an affine
Kac-Moody algebra $G_k$, where 
$k$ is the level and 
$G$ is a  semi-simple Lie group or supergroup.
As in the $SU(2)$ case,  one can use  
the  free field realization of these models \cite{W,SV}.
For the $G_k$-WZW model the singular limit is given   
by  $k + h \rightarrow 0 $, where $h$ is the dual
Coxeter number of $G$ \cite{FF}.
The charge and spin independent pairing
Hamiltonians studied by Richardson \cite{R6,R7}
in the context on Nuclear Physics 
probably  belong to this category of models.

{\bf Acknowledgments} I want to thank specially J. Dukelsky
for many discussions on the subject. I want also to acknowledge
conversations with  A. Belavin, E.H. Kim, 
M.A. Mart{\'i}n-Delgado,  A. Ramallo and  J. von Delft.  
This work was supported by the DGES spanish grant
PB97-1190.

{\bf Note added} After completion of this work we have been
informed by prof. A. Belavin about some related work by
H.M. Babujian \cite{Ba}, where he applies the Bethe ansatz
and the Knizhnik-Zamolodchikov eqs. to the Gaudin magnets \cite{Ga}.
The Gaudin's  model is given essentially by the $g \rightarrow 
\infty$ limit of the reduced BCS model. In fact the Gaudin's Hamiltonians
can be identified with $\lim_{g \rightarrow \infty } R_j/g$, where
$R_j$ are the CRS conserved quantities define in eq.(\ref{15}).


\begin{thebibliography}{99}
\bibitem{note} There is a sign difference between eq.(\ref{15})
and the one used by CRS in \cite{CRS}. This is due to our convenium
for $t^0_i$, which is opposite from that of CRS.
\end{thebibliography}
\end{document}